\documentclass[12pt]{iopart}

\usepackage{graphicx}
\usepackage{amssymb}

\newcommand{\eq}{Eq.~}
\newcommand{\eqs}{Eqs.~}
\newcommand{\fig}{Fig.~}
\newcommand{\figs}{Figs.~}
\newcommand{\cf} {cf.~}

\newcommand{\rref} {Ref.~}
\newcommand{\rrefs} {Refs.~}

\expandafter\let\csname equation*\endcsname\relax
\expandafter\let\csname endequation*\endcsname\relax
\usepackage{amsmath}
\begin{document}

\title{Quantum jump statistics with a shifted jump operator in a chiral waveguide}

\author{Dario Cilluffo,$^{1}$ Salvatore Lorenzo,$^{1}$ G. Massimo Palma,$^{1,2}$ and Francesco Ciccarello$^{1,2}$}

\address{\mbox{$^{1}$Universit\`a  degli Studi di Palermo, Dipartimento di Fisica e Chimica -- Emilio Segr\'e,}
\\\mbox{ via Archirafi 36, I-90123 Palermo, Italy}\\
	$^{2}$NEST, Istituto Nanoscienze-CNR, Piazza S. Silvestro 12, 56127 Pisa, Italy}

\ead{dario.cilluffo@unipa.it}

\vspace{10pt}
\begin{indented}
\item[]April 2019
\end{indented}

\begin{abstract}
Resonance fluorescence, consisting of light emission from an atom driven by a classical oscillating field, is well-known to yield a sub-Poissonian photon counting statistics. This occurs when only emitted light is detected, which corresponds to a master equation (ME) unraveling in terms of the canonical jump operator describing spontaneous decay. Formally, an alternative ME unraveling is possible in terms of a shifted jump operator. We show that this shift can result in sub-Poissonian, Poissonian or super-Poissonian quantum jump statistics. This is shown in terms of the Mandel Q parameter in the limit of long counting times, which is computed through large deviation theory. We present a waveguide-QED setup, comprising a chiral waveguide coupled to a driven atom, where photon counting is described by the considered class of shifted jump operators.

\end{abstract}

\section{Introduction}

A remarkable consequence of the 
wave function collapse is occurrence of {\it quantum trajectories} \cite{Carmichael2002open,GardinerQN00, PlenioRMP98, WisemanMilburnBook} when continuously monitoring the dynamics of small quantum systems such as a single atom. These are stochastic dynamics
that typically feature quantum jumps: sudden changes of the system's state occurring at unpredictable times. 
As a major achievement of modern quantum technologies, addressing small quantum systems is nowadays possible, which enables observation of quantum jumps and quantum trajectories in the lab \cite{gleyzesNAT2007,hatridgeARX2019}.

A quantum trajectory corresponds to a specific sequence of measurement outcomes on probes after that each probe has weakly and shortly interacted with the system under study (weak measurements) \cite{WisemanMilburnBook}.  Averaging over a large number of trajectories, namely measurement outcomes, results in a deterministic evolution of the density matrix governed by the celebrated Lindblad master equation (ME) \cite{breuerTheory2007}. Notably, the same ME can be obtained from different {detection schemes} ({\it unravelings}) corresponding to different types of weak measurements. 
Changing unraveling can deeply modify the nature of quantum trajectories, in particular the {\it counting statistics} of quantum jumps \cite{Carmichael2002open}, which is our focus in this work.

A paradigmatic process with non-trivial quantum-jump statistics is {\it resonance fluorescence}, a phenomenon of central importance in quantum optics \cite{Scully1997Quantum}: an atomic transition is resonantly driven by a coherent drive, causing the atom to cycle between the ground and excited states. When the atom is in the excited state, it may emit a photon via spontaneous emission. Recording the emitted photon by a detector projects the atom in the ground state, witnessing occurrence of a quantum jump. The resulting counting statistics, where each count corresponds to a detected photon namely a quantum jump on the atom, is known to be sub-Poissonian (showing the non-classical nature of the light produced in this process) \cite{Carmichael2002open,mandelOptLett1979}. This is essentially because, once the atom has released a photon by jumping to the ground state, it needs some time before it get excited again and emit a second photon. This gives rise to photon anti-bunching.

The standard quantum-trajectory description of resonance fluorescence is in terms of an unraveling with jump operator $\hat L=\sqrt{\gamma}\,\hat\sigma_-$ with $\gamma$ the spontaneous decay rate and $\hat\sigma_-$ the usual pseudo-spin ladder operator on the atom. This jump operator describes a photodetector recording only photons {\it emitted} from the atom. In free space, this is physically justified since the atom emits in all directions while the driving beam propagates along a well-defined direction. Accordingly, superposition effects between incoming and emitted light are negligible.
However, assuming that the light produced by spontaneous emission can be detected independently of the input signal is generally not possible in {one dimension} (1D), i.e., when incoming and emitted photons are constrained to propagate along the same direction. Such geometries can nowadays be implemented in a variety of setups, which typically comprises (pseudo) atoms -- even a single one -- efficiently coupled the 1D field of a waveguide \cite{RoyRMP17,LiaoPhyScr16,GuPhysRep2017}. In these settings, currently the focus of an emerging subfield of quantum optics called waveguide QED, dramatic interference effects between emitted and input light can take place. This can for instance lead to complete extinction of a propagating wave due to scattering from one artificial atom \cite{AstafievSci10}.

A detector unable to distinguish between emitted and input photons has a corresponding jump operator $\hat L=\sqrt{\gamma}\,\hat\sigma_-+i\alpha$, that is the usual jump operator accounting for spontaneous decay but {\it shifted} by a c-number proportional to the amplitude of the input coherent beam \cite{WisemanMilburnBook}. Shifted operators of this form appeared in recent works investigating waveguide-QED setups \cite{ManzoniNatComm17,ZhangPRA18,ZhangPRL19}. \rref\cite{ZhangPRA18}, in particular, studied quantum jump statistics in a two-way waveguide, in which case the above shifted jump operator is required for counting transmitted photons.

Notably, on a merely {\it formal} ground, the standard resonance fluorescence ME (corresponding to optical Bloch equations) can always be unraveled in terms of a shifted jump operator, provided that the Rabi frequency entering the drive Hamiltonian is suitably redefined \cite{WisemanMilburnBook}. 
While this shift obviously does not affect the unconditional dynamics, one can expect the quantum jump statistics to change: providing evidence of these changes is the goal of this paper. We specifically focus on the limit of {\it long counting times}, which enables prompt computation via large deviation theory \cite{garrahan2010thermodynamics} of the scaled cumulant generating function, hence all the moments of the photon counting distribution. We show that, depending on the parameters, the Mandel Q parameter can take on positive or negative values or vanish, thus exhibiting signatures of super-Poissonian, sub-Poissonian or Poissonian statistics. The considered unraveling with shifted jump operator -- which in free space does not correspond to realistic detection schemes (at least in an obvious way) -- is shown to describe the photon counting statistics in a chiral waveguide coupled to an atom \cite{LodahlReviewNature17}, which is driven at once by a classical field (external to the waveguide) and a coherent beam travelling along the waveguide.

This paper is organized as follows. In Section \ref{section-RF}, we briefly review basic notions of quantum trajectories and counting statistics in the specific context of standard resonance fluorescence. In Section \ref{section-shift}, we introduce the unraveling in terms of a shifted jump operator and discuss its implementation in a waveguide-QED setup. In Section \ref{section-LDT}, we briefly review the large-deviation-theory approach to work out all the moments of the photon counting distribution for long counting times. In Section \ref{section-results}, we investigate the quantum jump statistics as a function of the jump operator shift by analyzing the behavior of the Mandel Q parameter in some paradigmatic instances. Finally, we draw our conclusions in Section \ref{concl} .

\section{Quantum jump statistics in standard resonance fluorescence}\label{section-RF}

Consider a two-level atom, with ground (excited) state $|g\rangle$ ($|e\rangle$) and associated ladder operators $\hat \sigma_-=\hat\sigma_+^\dag=|g\rangle\langle e|$, which is driven by a classical oscillating field with Rabi frequency $\Omega$ and resonant with the atom. The atom is at the same time subject to spontaneous emission with decay rate $\gamma$.
In a frame rotating at the drive frequency, the atomic state $\rho$ evolves in time according to the Lindblad master equation (ME) \cite{gorini1976completely,lindblad1976generators}
\begin{equation}
\dot \rho=- i\, [\hat H,\rho]+\mathcal D({\hat L})\,\rho \,,\label{ME}
\end{equation}
(we set $\hbar=1$ throughout) with
\begin{align}
\mathcal D({\hat L})\,\rho=\hat L \,\rho\,\hat L^\dag-\tfrac{1}{2}\,  (\hat L^\dag\hat L \rho+\rho\,\hat L^\dag\hat L  )\,,
\end{align}
where the driving Hamiltonian $\hat H$ and jump operator $\hat L$ are respectively given by
\begin{eqnarray}
\qquad\qquad\hat H=\Omega \left(\hat\sigma_++\hat\sigma_-\right)\label{H}\,,\qquad\qquad\hat L=\sqrt{\gamma}\,\hat\sigma_-\,.
\label{HL}
\end{eqnarray}

ME \eqref{ME} can be "unraveled" in terms of \eqref{HL} by expressing its solution at time $t$ as \cite{Carmichael2002open}
\begin{align}
\rho_t=\mathcal{R}_t\rho_0+\sum_{K=1}^\infty \int_0^t dt_K\,\cdots \int_0^{t_2} dt_1\,\mathcal{R}_{t-t_K}\mathcal{J}\mathcal{R}_{t_K-t_{K-1}}\cdots \mathcal{J}\mathcal{R}_{t_2-t_1}\mathcal{J}\mathcal{R}_{t_1}\rho_0\,,\label{aver}
\end{align}
where we defined the superoperators
\begin{align}
\mathcal{R}_t \,\rho=e^{-i \left(\hat H-\tfrac{i}{2} \hat L^\dag \hat L\right) t}\rho \,e^{i \left(\hat H-\tfrac{i}{2} \hat L^\dag \hat L\right) t}\,,\,\,\,\,\,\mathcal{J}\rho=\hat L\, \rho\, \hat L^\dag\,\,.\label{so}
\end{align}
A specific photon counting experiment consists of a quantum trajectory during which an emitted photon is recorded at times $t_1,t_2,...,t_K$. At these times, the atom undergoes a quantum jump to the ground state. Between two next jumps, the atom evolves according to the non-Hermitian Hamiltonian $\hat H_{\rm eff}=\hat H-\tfrac{i}{2} \hat L^\dag \hat L$. The deterministic (unconditional) dynamics described by ME \eqref{ME} arises by averaging over a large number of photon counting experiments.

The number of jumps $K$ in a trajectory is a random variable. 
The counting statistics of jumps/photon counts is defined by the knowledge of ${P_t(K)}$ for any $K$, where ${P_t(K)}$ is the probability to get $K$ counts within the time window $[0,t]$ with $t$ the detector's {counting time}. Statistical moments are given by $\langle K^n \rangle=\sum_{K=0}^\infty P_t(K) K^n$, with the first and second moments in particular allowing to determine
\begin{eqnarray}
\qquad \qquad k=\frac{\langle K \rangle}{t}\,,\qquad\qquad Q=\frac{\langle K^2 \rangle-\langle K \rangle^2}{\langle K \rangle}-1\label{k0Q}\,,
\end{eqnarray}
which are the average number of counts per unit time or activity, and the Mandel Q parameter, respectively. The latter is especially important since it provides informations about correlations of emitted photons. When $Q=0$, the counting statistics is Poissonian [$P_t(K)$ is a Poisson distribution] witnessing no correlations between emitted photons. Instead, $Q>0$ and $Q<0$ respectively correspond to a super- and sub-Poissonian statistics the latter having no classical counterpart. Occurrence of super-Poissonian statistics typically occurs when photons tend to bunch together (photon bunching). Sub-Poissonian statistics is instead a signature of photon anti-bunching, hence jumps tend to occur at well-separate times \cite{LoudonQTL03}.

In the limit of long counting times, $t\rightarrow \infty$, the activity and Mandel parameter for unraveling \eqref{unraveling} are respectively given by \cite{mandelOptLett1979} $k=4  \gamma  \Omega^2/({8 \Omega^2{+}\gamma^2})$ and 
\begin{equation}
Q=-\frac{24 \gamma^2\Omega^2  }{(8 \Omega^2+\gamma^2 )^2}\,.\label{Qst}
\end{equation}
Function Q is always negative indicating occurrence of sub-Poissonian statistics and photon anti-bunching of the emitted light.

\section{Shifted-jump-operator unraveling and physical implementation}\label{section-shift}

As is well-known, there are infinite sets of effective Hamiltonian and jump operators in terms of which a given ME can be expressed. Each set defines a ME unraveling and generally gives rise to a  different statistics of quantum jumps.

A class of unravelings for the resonance fluorescence ME \eqref{HL} is defined by
\begin{equation}
\hat H=\left(\Omega-\tfrac{1}{2}\sqrt{\gamma}\,\alpha\right)\hat\sigma_++{\rm H.c.} \,,\,\,\,\,\hat L=\sqrt{\gamma} \,\hat\sigma_- + i \alpha\,\,,\label{unraveling}
\end{equation}
where $\alpha$ is complex (there is only one jump operator). This is a  special case of the general property \cite{WisemanMilburnBook} according to which any Lindblad ME \eqref{ME} $ {\dot\rho}=- i [\hat H,\rho]+ \sum_\nu \mathcal D({\hat L_\nu})\,\rho$ is invariant under the transformation $\hat L_\nu\rightarrow \hat L_\nu+\chi_\nu$, $\hat H\rightarrow \hat H- \frac{i}{2} \sum_\nu (\chi_\nu^* \hat L_\nu- {\rm H.c.})$ for an arbitrary set of complex numbers $\{\chi_\nu\}$, which can be easily checked.
\begin{figure}
\centering
	\includegraphics[width=10.5cm]{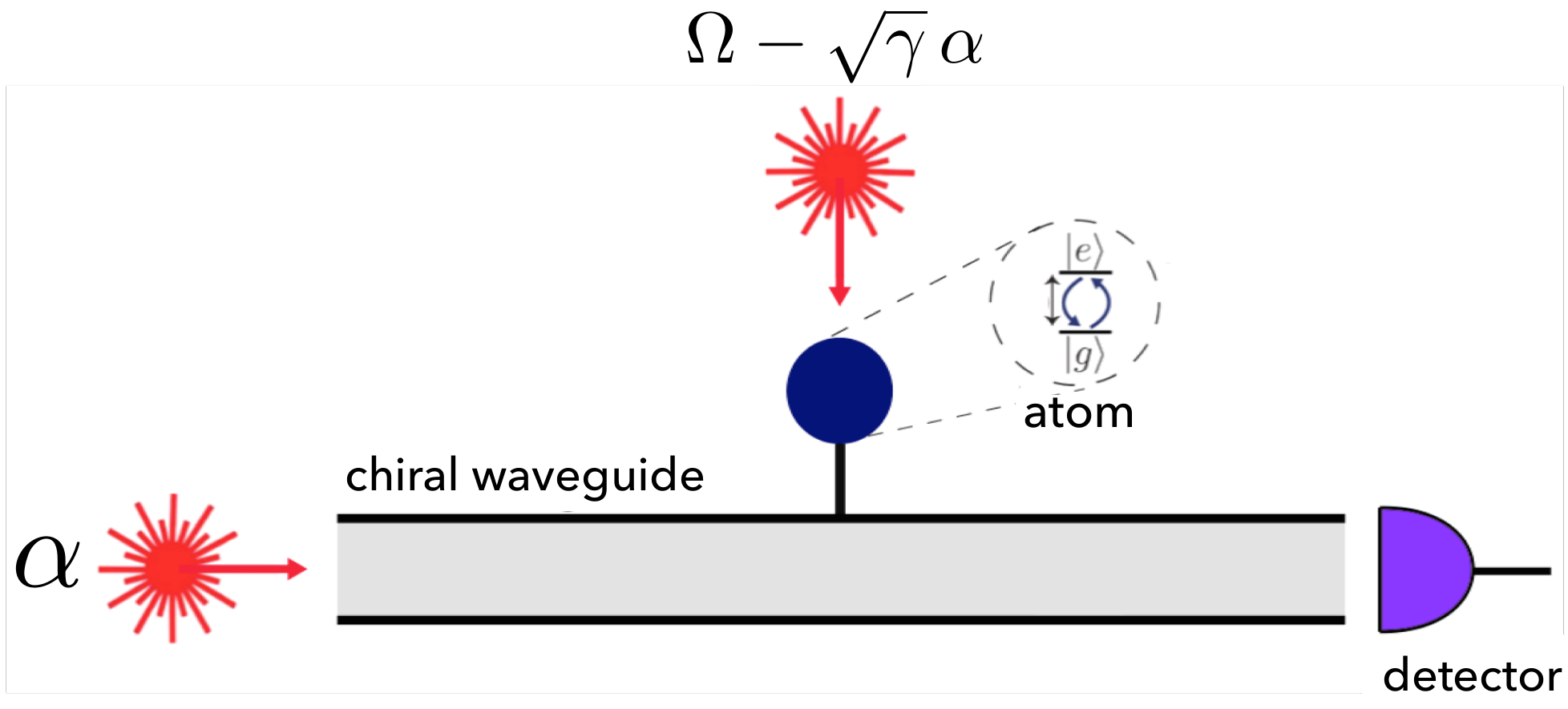}
	\caption{Waveguide-QED setup to implement unraveling \eqref{unraveling} with shifted jump operator. A two-level atom is coupled to a chiral waveguide, sustaining a one-dimensional field that propagates in one direction only. The atom is driven by two continuous-wave coherent beams: one with Rabi frequency $\Omega-\sqrt{\gamma}\,\alpha$ propagating out of the waveguide and another one of amplitude $\alpha$ traveling along of the waveguide. Photons are counted at one end of the waveguide: the detector senses light emitted from the atom {superposed} with the coherent wave $\alpha$. \label{fig1}}
\end{figure}
The ME unraveling \eqref{unraveling} consists in applying a c-number {\it shift} to the jump operator, which leaves ME \eqref{HL} invariant provided that the Rabi frequency in the Hamiltonian term is corrected as $\Omega\rightarrow \Omega-\tfrac{1}{2}\sqrt{\gamma}\,\alpha$. Shifted jump operators appeared for instance in \rrefs\cite{ManzoniNatComm17,ZhangPRA18,ZhangPRL19,CarmichaelPRL93}. Note that the replacement $\sqrt{\gamma} \,\hat\sigma_-\rightarrow \sqrt{\gamma} \,\hat\sigma_- + i \alpha$ tends to smooth out the effect of jumps in the sense that, when $\alpha$ is large, a jump acts trivially on the system leaving its state unchanged (after normalization).

We illustrate next a setup where photon counting statistics is described by unraveling \eqref{unraveling}. To this aim, consider first a chiral waveguide (enabling light propagation only in one direction) side-coupled with strength $\sqrt{\gamma}$ to a two-level atom under the rotating wave approximation (the waveguide has linear dispersion law). The setting is sketched in \fig\ref{fig1}. A coherent, continuous-wave beam of amplitude $\alpha$ resonant with the atom is injected at one end of the waveguide and scatters from the emitter. Note that there is no back-scattered light since the waveguide is chiral. Photons are counted at a detector placed beyond the atom. Applying input-output formalism, the output field sensed by the detector is given by $\hat a_{\rm out}(t)=-i\,\sqrt{\gamma}\,\hat\sigma_-(t)+\alpha$ \cite{WisemanMilburnBook}, indicating that it results from the coherent superposition of the incoming coherent beam and emitted light. Accordingly, the jump operator describing photon detection coincides with $\hat L$ in \eq\eqref{unraveling} in a way that the corresponding unraveling is defined by \cite{ZhangPRA18}
\begin{equation}
\hat H=\tfrac{1}{2}\sqrt{\gamma}\,\alpha\,\hat\sigma_++{\rm H.c.} \,,\,\,\,\,\hat L=\sqrt{\gamma} \,\hat\sigma_- + i \alpha\,\,,\label{unraveling2}
\end{equation}
where we absorbed a $1/\sqrt{2\pi}$ factor into the definition of $\alpha$.
Note that, in light of the ME invariance property discussed above, \eqref{unraveling2} gives rise to same {\it unconditional} dynamics one would get with $\hat H=\sqrt{\gamma}\,\alpha\,\hat\sigma_++{\rm H.c.}$ and $\hat L=\sqrt{\gamma} \,\hat\sigma_-$

Consider now the more general case that, in addition to the coherent beam along the waveguide, a second classical drive (external to the waveguide) is applied on the atom as shown in \fig\ref{fig1}. If this extra drive has Rabi frequency $\Omega-\sqrt{\gamma}\alpha$, then the Hamiltonian in \eqref{unraveling2} acquires an additional term $(\Omega-\sqrt{\gamma}\alpha) \hat\sigma_++{\rm H.c.}$ while the jump operator is unaffected. This yields precisely unraveling \eqref{unraveling}. 
In practice, as suggested by \fig\ref{fig1}, the classical drive is embodied by a second coherent beam involving electromagnetic modes not supported by the waveguide (atomic leakage into these modes is assumed to be negligible). Note that emitted light superposes only with the input beam traveling along the waveguide (before being recorded at the detector). 

The setup in \fig1 is similar to one recently investigated in \rref\cite{trivediPRB2018} (see \fig4 therein), which however focused on an external drive with finite duration.

\section{Photon counting statistics via large deviation theory}\label{section-LDT}

In the limit of long counting times of concern here, the full counting statistics of quantum jumps can be worked out without the need for Monte Carlo simulations by resorting to {\it large deviation theory} (LDT) \cite{touchette2009large}. This statistical mechanics tool has been adapted to quantum trajectories in recent years \cite{garrahan2010thermodynamics,garrahanPhysA2018} and used in some quantum optics problems \cite{garrahan2011quantum,GarrahanPRA2018, Flindt2019, Alejandro2019}. 

The discrete probability distribution $P_t(K)$ can be equivalently represented through the associated moment-generating function
\begin{equation}
Z_t(s) =\langle e^{-s K}\rangle= \sum_{K=0}^{\infty} P_t(K) \,e^{-sK}\,, 
\label{partition}
\end{equation}
where, unlike $K$, $s$ is a continuous variable that can take values over the entire real axis. When the detector's counting time $t$ is long enough, $Z_t(s)$ takes the (so called) large-deviation form
\begin{equation}
Z_t(s)\sim e^{t\theta(s)}\,\,\,\,{\rm for \,\, long}\,\,t\,,\label{LDT}
\end{equation}
where $\theta(s)={\rm lim}_{t\rightarrow\infty}(1/t)\ln Z_t(s)$ is the scaled cumulant generating function (SCGF), at times referred to in the literature as LD function. Note that $\theta(0)=0$ since $Z_t(0)=\sum_K P_t(K){=}1$. The knowledge of the SCGF allows to work out all the cumulants, hence all the moments, of the counting distribution $P_t(K)$ through derivatives of $\theta(s)$ at $s{=}0$.
In particular, the activity and Mandel Q parameter (\cf equation \eqref{k0Q}) are given by
\begin{equation}
k=-\,\theta'(0)\,,\,\,\,Q=-\frac{\theta''(0)}{\theta'(0)} -1 \label{k0Q-2}
\end{equation}
It can be shown \cite{garrahan2010thermodynamics} that, for an unraveling defined by Hamiltonian $\hat H$ and a jump operators $\hat L$ (\cf\eq\eqref{ME}),
$\theta(s)$ is the largest real eigenvalue of the superoperator
\begin{equation}
\mathcal W_s (\rho)=- i [\hat H,\rho]+e^{-s}\hat L \,\rho\,\hat L^\dag-\tfrac{1}{2}\,  (\hat L^\dag\hat L \rho+\rho\,\hat L^\dag\hat L  )\label{Ws}\,,
\end{equation}
which for $s=0$ reduces to the canonical Lindblad-ME generator (right-hand side of equation \eqref{ME}). Thus, when the dimension of the open system is small as in the present case of a two-level atom, the counting statistics can in fact be worked out by solving a simple eigenvalue problem.
Clearly, turning the standard jump operator $\hat L=\sqrt{\gamma}\,\hat\sigma_-$ into the shifted one (see \eq\eqref{unraveling}) affects \eqref{Ws} in a non-trivial way. 

\subsection{Typical and rare trajectories}

One can define {\it biased} photocount probabilities as
\begin{equation}
P_t^{(s)} (K) = \frac{P_t (K) e^{-sK}}{Z_t(s)}\,,\label{Pbias}
\end{equation}
such that $P_t (K){=}P_t^{(s=0)} (K)$, and correspondingly biased moments $\langle K^n\rangle_s{=}\sum_{K} K^n P_t^{(s)}(K)$. 
Accordingly, the biased activity and Mandel parameter are  
\begin{equation}
k(s)={\langle K \rangle}_s/{t}\,,\,\,\,Q(s)=({\langle K^2 \rangle_s-\langle K \rangle_s^2})/{\langle K \rangle_s}-1\,,\label{Qs}
\end{equation}
which reduce to \eqref{k0Q} for $s=0$. It turns out that biased activity and Mandel parameter are related to the SCGF as
\begin{equation}
k(s)=-\,\theta'(s)\,,\,\,\,Q(s)=-\frac{\theta''(s)}{\theta'(s)}-1\label{k0Q-3}\,,
\end{equation}
which naturally extends \eqref{k0Q-2} to the entire $s$-axis. One can see that $k(s)<k(0)$ for $s>0$, while $k(s)>k(0)$ for $s<0$. Based on the above, the point $s=0$ corresponds to typical quantum trajectories, while non-zero values of $s$ to atypical trajectories. Trajectories for $s>0$ and $s<0$ are, respectively, less and more active than typical (since $k(s)$ is respectively lower and higher than $k(0)$) \cite{garrahan2010thermodynamics}.

\section{Mandel Q parameter for shifted jump operator}\label{section-results}

In \fig2 we set $\gamma=1$ and show the behavior of the $s$-dependent Mandel Q parameter \eqref{k0Q-3} as a function of $\alpha$ for $\Omega=0.1$ (a), $\Omega=0.25$ (b), $\Omega=1$ (c) and $\Omega=5$ (d).
\begin{figure}
\centering
	\includegraphics[width=17cm]{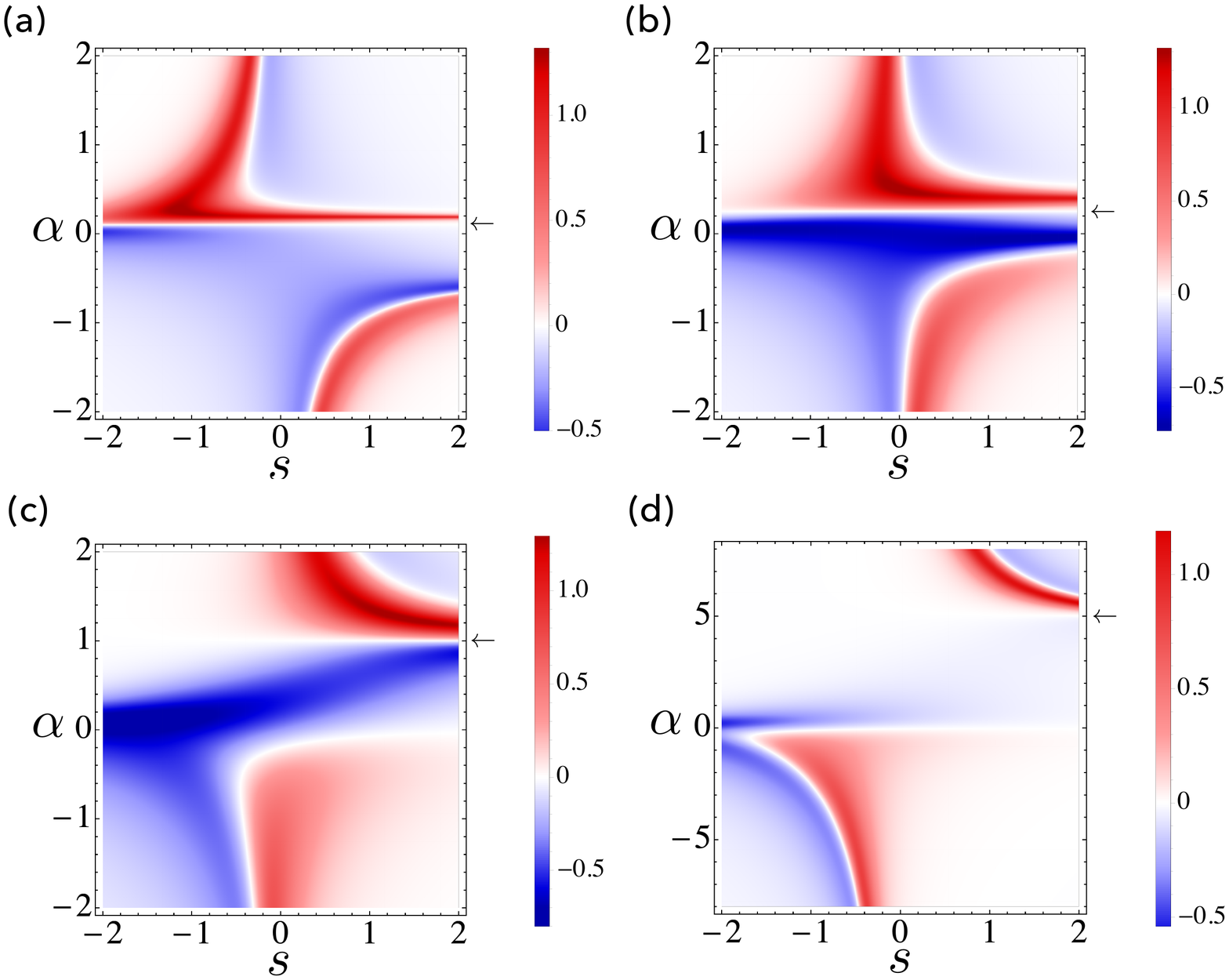}
	\caption{Mandel Q parameter $Q(s)$ as a function of $\alpha$ for $\Omega=0.1$ (a), $\Omega=0.25$ (b), $\Omega=1$ (c) and $\Omega=5$ (d) (we set $\gamma=1$). Each curve was obtained through a mesh of the $s$-axis and numerical diagonalization of \eqref{Ws}, from which $\theta(s)$ was inferred by selecting the largest eigenvalue. The $s$-dependent Mandel Q parameter $Q(s)$ was determined using \eqs\eqref{k0Q-3} after applying the finite difference method to compute the derivatives of $\theta(s)$. \label{fig2}}
\end{figure}
For $\alpha=0$ (standard unraveling), $Q$ generally grows with $s$, passing from negative values (sub-Poissonian statistics) to zero (Poissonian). In particular, $Q(0)$ (the canonical Mandel parameter for typical trajectories) is always negative and asymptotically approaches zero for growing Rabi frequency $\Omega$ in agreement with \eq\eqref{Qst}. 

For $\alpha\neq0$, i.e., under the shift of the jump operator (\cf\eq\eqref{unraveling}), the statistics is significantly affected and shows a rich behavior including occurrence of positive $Q$'s (super-Poissonian statistics). Note in particular that, for the considered values of $\Omega$, $Q(0)$ can be negative, positive or vanish by tuning $\alpha$.
For each set $\Omega$, $Q(s)=0$ identically for $\alpha=\Omega/\sqrt{\gamma}$ (horizontal white line marked with a little arrow on each panel). Above and below this line, different behaviors occur.

{\it Above the $\alpha{=}\Omega/\sqrt{\gamma}$ line}, in the case of \figs2(a) and (b), $Q(s)$ takes a positive (super-Poissonian) maximum followed by a negative (sub-Poissonian) minimum, exhibiting a steep drop in the passage from positive to negative values, witnessed by the occurrence of a sharp white curve. This curve bends so as to become asymptotically parallel, for $s$ large enough, to the $\alpha=\Omega/\sqrt{\gamma}$ line. Hence, above the latter a super-Poissonian "band" arises. In the case of \figs2(c) and (d), a somewhat similar behavior occurs but at farther values from $s=0$, and the sharp super-Poissonian band is no longer present unless $s$ is very large.

{\it Below the $\alpha{=}\Omega/\sqrt{\gamma}$ line}, a region of predominantly sub-Poissonian statistics appears (see \figs2(a)-(d)), which becomes more and more Poissonian for large $\Omega$'s (see \fig2(d)). Below this sub-Poissonian region, a behavior occurs that consists in a sharp transition from sub-Poissonian to super-Poissonian statistics as $s$ grows from negative to positive values. The transition sharpness is witnessed by a white curve of Poissonian statistics clearly visible on each panel.

We in particular highlight the outcomes in \fig2(b) corresponding to $\Omega=\gamma/4$. Here, in addition to $\alpha=\Omega/\sqrt{\gamma}$ where as discussed $Q(s){=}0$ identically, another $s$-independent behavior occurs for $\alpha=0$ in which case $Q(s){=}-2/3$, which retrieves what was found in \rrefs\cite{garrahan2010thermodynamics,garrahanPhysA2018} for standard unraveling. Also, note that large portions of the two Poissonian curves (along which $Q(s)$ vanishes) run very close to $s=0$ resulting in two crossovers. Indeed, for $\alpha\gtrsim \gamma$, a pronounced crossover from super-Poissonian to sub-Poissonian statistics occurs when passing from trajectories more active than typical ($s<0$) to less active ones ($s>0$). For $\alpha\lesssim -\gamma$, another crossover takes place but now from sub-Poissonian to super-Poissonian statistics.

The Poissonian white horizontal line in each panel can be explained by referring to \fig1 as follows. Indeed, for $\alpha=\Omega/\sqrt{\gamma}$, the external drive amplitude $\Omega-\!\sqrt{\gamma}\,\alpha$ vanishes leaving the atom exposed only to the coherent beam traveling along the waveguide (thereby in this case \eqref{unraveling} coincides with \eqref{unraveling2}). Since the waveguide is chiral and we work in the limit of infinite counting time, in each trajectory the number of photons injected into the waveguide exactly matches the number of photons detected at the waveguide's end (see \fig1). Since the incoming photons are Poisson-distributed (the beam is a coherent state), so will be the detected ones. Accordingly, the Mandel Q parameter identically vanishes. This is generally no longer the case when the external drive is also present ($\alpha\neq\Omega/\sqrt{\gamma}$): photons from the external beam that are not absorbed by the atom cannot reach the detector. In this case, the detector counts photons emitted by the atom - which alone would give rise to sub-Poissonian statistics - coherently superposed with Poisson-distributed photons of the beam traveling along the waveguide. The resulting interference yields the complex behavior in \fig2, including occurrence of super-Poissonian counting statistics in some regions of the parameters space.

\section{Conclusions} \label{concl}

In summary, we presented a study of the effects of a shifted-jump unravelling on the photon counting statistics in resonance fluorescence. The standard unraveling assumes that only light emitted from the atom is detected, the corresponding jump operator thus being proportional to the usual annihilation spin ladder operator. In a waveguide geometry, this may not be the case so that the jump operator can feature a c-number shift. For fixed Rabi frequency $\Omega$, this shift does not change the unconditional dynamics of the atom (master equation) but quantum trajectories are generally affected in a non-trivial way. We have shown this by focusing on the counting statistics of quantum jumps in the limit of infinite counting time. This limit allows to compute quickly all the moments of the photon counting distribution using large deviation theory. We in particular explored the effect of the jump-operator shift on the behavior of the $s$-dependent Mandel Q parameter (with $s$ the  continuous variable conjugated to the number of photon counts). This was shown to exhibit a rich behavior, including occurrence of all the three statistics (sub- and super-Poissonian and Poissonian) as opposed to the well-known sub-Poissonian statistics characteristic of standard resonance fluorescence. While the jump-operator shift can be conceived {\it formally} as a well-defined alternative unraveling of the standard resonance fluorescence master equation,  in free space it does not correspond to realistic detection schemes. We thus illustrated a waveguide-geometry setup where it naturally describes photon counting experiments \footnote{Photon counting statistics in waveguide-QED setups was studied also through different approaches (not based on quantum trajectories), see e.g. \cite{ChumakPRA2013,PletyukhovPRA2015}}. This comprises a chiral waveguide side-coupled to an atom driven at once by a coherent beam propagating along the waveguide and one external to it. This implementation helps interpreting the counting statistics as the result of non-trivial interference between sub-Poissonian-distributed light emitted from the atom and Poissonian light from the beam traveling along the waveguide.

The material presented in this short paper should be intended as a preliminary investigation of photon counting statistics in waveguide geometries featuring emitters driven at once by drives propagating along different directions. Further insight into the observed behaviors, for instance to clarify the physical origin of the second zero of $Q(0)$ as a function of $\alpha$ in \figs2(a)-(d), requires a more detailed analysis in terms of quantities such as the second-order correlation function or the waiting time distribution that will appear elsewhere.

\section*{Acknowledgments}

We acknowledge fruitful discussions with H. U. Baranger, G. Buonaiuto, I. Lesanovsky, B. Olmos, R. Trivedi, and X. H. H. Zhang.
We acknowledge support under PRIN project 2017SRN- BRK QUSHIP funded by MIUR.
\bigskip
\bigskip

\bibliographystyle{ieeetr}			
\bibliography{WQED}

\end{document}